\documentclass[aps,pra,amssymb,amsmath,superscriptaddress,twocolumn,10pt]{revtex4-2}
\usepackage{xcolor}
\usepackage{graphicx}
\usepackage{bm}
\usepackage{times,txfonts}
\usepackage[english]{babel}
\usepackage{hyperref}
\usepackage{booktabs}
\usepackage[normalem]{ulem}
\usepackage{soul}
\usepackage{subfigure}

\def\ee{{\mathrm{e}}}
\def\ii{{\mathrm{i}}}
\def\dd{{\mathrm{d}}}

\def\bra#1{\langle #1 |}
\def\ket#1{| #1 \rangle}

\begin{document}

\title{The dressed atom revisited: Hamiltonian-independent treatment of the radiative cascade}

\author{Francesco V. Pepe}
\affiliation{Dipartimento Interateneo di Fisica, Universit\`{a} degli Studi di Bari, I-70126 Bari, Italy}
\affiliation{Istituto Nazionale di Fisica Nucleare, Sezione di Bari, I-70125 Bari, Italy}

\author{Karolina S{\l}owik}
\affiliation{Institute of Physics, Nicolaus Copernicus University in Toru\'n, Grudziadzka 5, Toru\'n 87-100, Poland}

\begin{abstract}
    The dressed atom approach provides a tool to investigate the dynamics of an atom-laser system by fully retaining the quantum nature of the coherent mode. In its standard derivation, the internal atom-laser evolution is described within the rotating-wave approximation, which determines a doublet structure of the spectrum and the peculiar fluorescence triplet in the steady state. 
    However, the rotating wave approximation may fail to apply to atomic systems subject to femtosecond light pulses, light-matter systems in the strong-coupling regime or sustaining permanent dipole moments.
    This work aims to demonstrate how the general features of the steady-state radiative cascade are affected by the interaction of the dressed atom with propagating radiation modes. Rather than focusing on a specific model, we analyze how these features depend on the parameters characterizing the dressed eigenstates in arbitrary atom-laser dynamics, given that a set of general hypotheses is satisfied.
    Our findings clarify the general conditions in which a description of the radiative cascade in terms of transition between dressed states is self-consistent. We provide a guideline to determine the properties of photon emission for any atom-laser interaction model, which can be particularly relevant when the model should be tailored to enhance a specific line.
    We apply the general results to the case in which a permanent dipole moment is a source of low-energy emission, whose frequency is of the order of the Rabi coupling.
\end{abstract}

\maketitle
The dressed atom approach is a tool to investigate the dynamics of an atom interacting with a laser mode in a fully quantum way, along with its coupling to the propagating radiation modes \cite{cohentannoudji1977dressed}. Compared to other powerful methods to investigate the atom-laser interactions such as the optical Bloch equations \cite{allen1975optical,mollow1969power}, the dressed-atom approach has various advantages \cite{cohentannoudjiAPI}. First, the atom-laser dynamics is typically described by time-independent evolution equations, either in a closed-system or in an open-system framework; thus, by avoiding at all the introduction of a classical time-dependent field, one can potentially extend the analysis well beyond the rotating-wave approximation (RWA), where time dependence can hardly be integrated out of the Bloch equations. Second, the quantum nature of the laser mode allows for a consistent description in both the time and frequency domains \cite{cohentannoudjiAPI}, as well as for the formulation of a decay dynamics that properly takes into account hybridization of the unstable atom state and non-perturbative energy shifts induced by the atom-laser coupling. The dressed-atom approach was originally introduced to describe \textit{natural} systems interacting with light, in a condition where a quasi-resonant laser and weak coupling made RWA safely applicable. However, recent developments in the implementation of \textit{artificial} atoms allowed for the realization of strong light-matter coupling \cite{englund2007controlling,sornborger2004superconducting,leibfried2003quantum,wallraff2004strong,forndiaz2010observation,niemczyk2010}. Especially in view of their relevance for the implementation of quantum protocols \cite{pellizzari1995decoherence}, strong-coupling effects call for a specific theoretical focus on light-matter interaction beyond RWA \cite{braak2011}.

The purpose of this Letter is to present a general framework describing the dynamics of a dressed two-level atom when one of the atomic states is unstable, yielding a radiative cascade in the asymptotic regime \cite{cohentannoudjiAPI}. While the dressed-atom approach in its standard form is based on the RWA (i.e., the Jaynes-Cummings model \cite{shore1993}), and can be potentially corrected by perturbing the system with additional terms, we do not aim at a formulation tailored on a specific Hamiltonian model of atom-laser interactions. Instead, we show how the general features of the radiative cascade depend on the parameters that characterize the dressed atom-laser eigenstates, given a set of minimally restrictive phenomenological hypotheses on the system dynamics. The general results can then be applied to a variety of cases, including, without claiming to be exhaustive, the presence of a permanent dipole moment for one or both the atomic levels \cite{kibis2009,chestnov2017,scala2021beyond}, the non-negligible contribution of counter-rotating terms \cite{xie2017,niemczyk2010,crespi2012,schneeweiss2018,braak2011,chen2012}, the presence of nonlinear coupling in the field operators \cite{duan2022unified}.
We will apply the theory to the specific case systems with permanent dipole moments, that describe, e.g., polar molecules, or engineered condensed-matter systems such as asymmetric quantum dots. Their optical response enriched by the interplay of permanent and induced electric dipoles leads to application potentials for THz radiation sources \cite{kibis2009,paspalakis2013,chestnov2017,gladysz2020}, squeezed light generation \cite{koppenhofer2016,anton2017} and for nonlinear optical absorption \cite{paspalakis2013}. Their spontaneous decay dynamics  was analyzed in Ref.~\cite{savenko2012} and, with the addition of counter-rotating terms, in Ref.~\cite{scala2021beyond}. 

The results here derived are based on the following set of hypotheses:
\begin{enumerate}
    \item The atom can be treated as a two-level system, whose Hilbert space is spanned by the states $\{\ket{g},\ket{e}\}$. While $\ket{g}$ is stable, in the absence of coupling with the laser the state $\ket{e}$ spontaneously decays towards $\ket{g}$ by single-photon emissions in a non-dispersive medium, with a decay rate $\Gamma_0$. Such an atom-photon interaction can be treated in the Markovian approximation.
    \item The average $\langle N \rangle$ and the standard deviation $\Delta N =\sqrt{ \langle (N - \langle N \rangle)^2 \rangle }$ of the photon number in the laser mode satisfy 
    \begin{equation}\label{eq:laserstat}
        \langle N \rangle \gg \Delta N \gg 1 .
    \end{equation}
    This assumption is crucial to approximate the action of the mode annihilation operator $a$ on number states $\ket{N}$ as $a\ket{N}\simeq \sqrt{\langle N \rangle} \ket{N-1}$ and obtain atom-laser transition matrix elements that are constant for all practical purposes. Moreover, the system is observed for a time $T$ such that the decay-induced depletion does not significantly alter the laser number statistics, i.e.\ $\Gamma_0 T\ll \Delta N$.
    \item The energy diagram of the atom-laser system is made of non-degenerate doublets of \textit{dressed eigenstates} $\{\ket{1(N)},\ket{2(N)}\}$, with eigenvalues
    \begin{equation}
        E_j(N) = \hbar \left( N\omega_L - (-1)^{j}\frac{\Omega}{2} \right) \,\,\, \text{for } j\in\{1,2\} ,
    \end{equation}
    where $\Omega<\omega_L$, at least in the region $N\simeq \langle N \rangle$ that is relevant for dynamics.
    \item The inverse time scale of the decay processes is the smallest one in the system dynamics, satisfying
    \begin{eqnarray}\label{eq:enscales}
        \omega_L,\Omega,\omega_L-\Omega \gg \Gamma_0 .
    \end{eqnarray}
    This property allows neglecting interference between emission lines centered on the three frequencies at the left-hand side, and \textit{a fortiori} on higher ones, which is equivalent to applying the secular approximation in the standard treatment of the radiative cascade \cite{cohentannoudjiAPI}.
\end{enumerate}

\begin{figure}
    \centering 
    \includegraphics[width=0.4\textwidth]{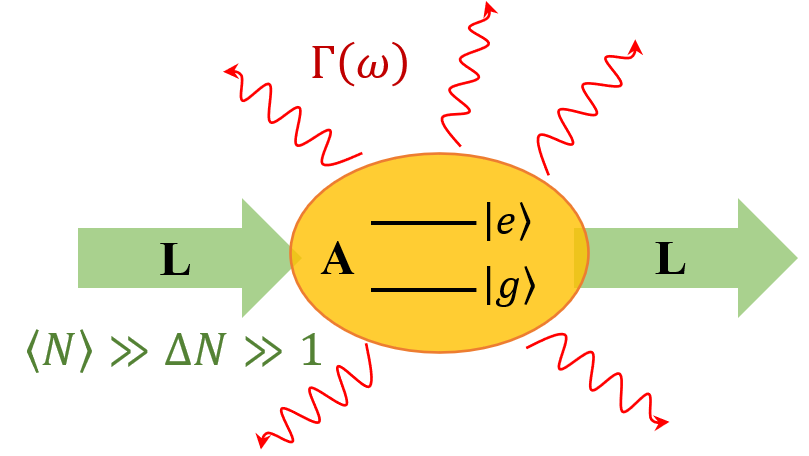} \\ \vspace{.5cm}
    \includegraphics[width=0.4\textwidth]{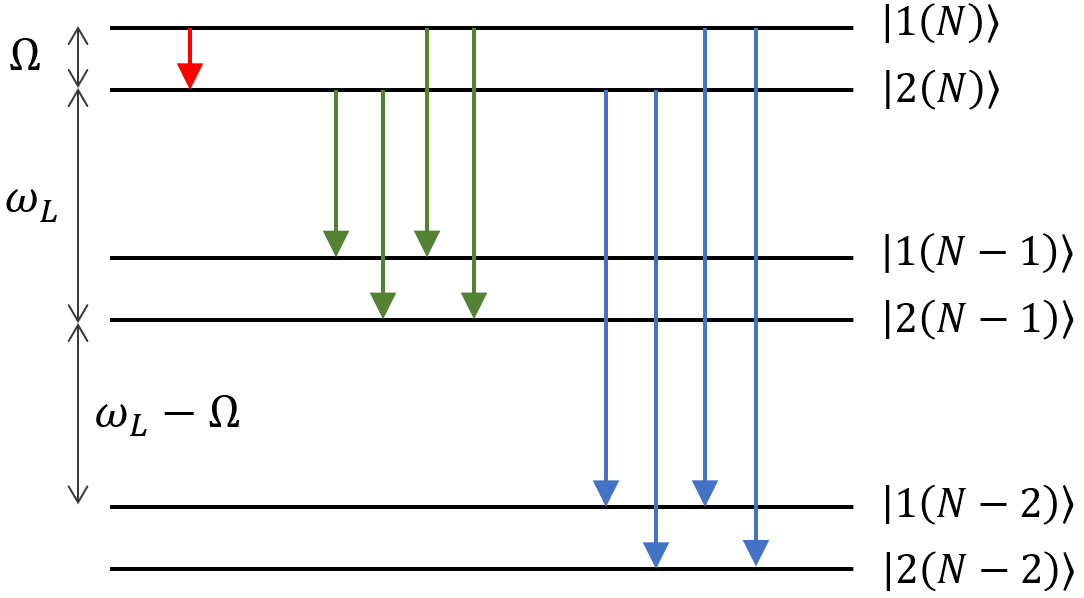}
    \caption{\textit{Upper panel.} Schematic representation of a two-level atom coupled both to a stable laser mode, and to radiation modes characterized by a continuous spectrum, weighed at different frequencies by the form factor $\Gamma(\omega)$.
    \textit{Lower panel.} Doublet structure of the spectrum of the atom-laser Hamiltonian $H_{\mathrm{AL}}$ and possible transitions between them, due to the contributions of the atomic excited and ground states in the eigenstates. The elementary frequency scales characterizing the spectrum are all much larger than the excited atom decay rate.}
    \label{fig:transitions}
\end{figure}
Give the above conditions, the whole system dynamics, including the radiation field modes, can be expressed by the Hamiltonian $H=H_{\mathrm{AL}}+H_{\mathrm{R}}+V_{\mathrm{AR}}$, where
\begin{align}
    H_{\mathrm{AL}} & = \sum_{j=1,2}\sum_N E_j(N) \ket{j(N)}\bra{j(N)} , \\
    H_{\mathrm{R}} & = \hbar \int \dd\omega\, \omega a^{\dagger}(\omega) a(\omega) 
\end{align}
respectively determine the dynamics of the atom-laser system and the free radiation field, represented by the operators $a(\omega)$ that satisfy canonical commutation relations $[a(\omega),a^{\dagger}(\omega')]=\delta(\omega-\omega')$. Note the general form of the atom-light Hamiltonian $H_\mathrm{AL}$, for which we do not assume specific form of the interaction.
Finally,
\begin{equation}\label{eq:VAR}
    V_{\mathrm{AR}} = \hbar \int \dd\omega \, \sqrt{\frac{\Gamma(\omega)}{2\pi}} \left ( \mathcal{S}_+ a(\omega) + \mathcal{S}_- a^{\dagger}(\omega) \right) ,
\end{equation}
with $\mathcal{S}_+=\ket{e}\bra{g}$ and $\mathcal{S}_-=\ket{g}\bra{e}$, describes the RWA coupling of the atom with the continuum field modes, with an arbitrary form factor $\Gamma(\omega)$ vanishing for $\omega\leq 0$ that determines the properties of the continuum. The expressions of the transition operators $\mathcal{S}_{\pm}$ in the dressed eigenstate basis are determined by the coefficients (see Appendix)
\begin{equation}\label{eq:Aq}
    A_{ij}^{(q)} = \sum_p \langle i(N+p+q)|e,N \rangle \langle g,N|j(N+p) \rangle ,
\end{equation}
with $\{\ket{e,N},\ket{g,N}\}$ being the simultaneous eigenstates of the free atom Hamiltonian and of laser photon number, can be assumed $N$-independent due to the hypothesis \eqref{eq:laserstat} on the laser state. The form of the term $V_{AR}$, where all the irrelevant field degrees of freedom have been integrated out, contains all the relevant information to determine, using Fermi's golden rule, the decay rate $\Gamma_0=\Gamma(\omega_0)$ of the state $\ket{e}$ when the laser is off, with $\omega_0$ the bare transition frequency between the two atomic states \cite{scala2020light}. Neglecting the frequency dependence of the form factor is a common \textit{flat coupling} approximation \cite{cohentannoudjiAPI}, which we show below may lead to incorrect results beyond the RWA regime.

The frequency scale hierarchy allows the formulation of Markovian dynamics of the atom-laser system by separately considering dissipative processes associated with transitions at different frequencies. We assume that the field continuum starts in its vacuum state. Since no restriction is required on the composition of the dressed eigenstates in terms of the atomic states $\ket{e}$ and $\ket{g}$, the dissipative part of the Gorini-Kossakowski-Lindblad-Sudarshan equation reads
\begin{align}\label{eq:Ldiss}
    \mathcal{L}_{\mathrm{diss}}(\sigma) = & \sum_{q,N,N'} \Bigl[ \sum_{i,j=1,2} \Gamma_{ij}^{(q)} \Bigl( - \frac{\delta_{N,N'}}{2}  \bigl\{ \sigma, \ket{i(N)}\bra{i(N)} \bigr\}   \nonumber \\
    & + \sigma_{i(N+q),i(N'+q)} \ket{j(N)}\bra{j(N')} \Bigr) \nonumber \\
    & + K_{12}^{(q)} \sigma_{1(N+q),2(N'+q)} \ket{1(N)}\bra{2(N')} \nonumber \\
    & + \bigl(K_{12}^{(q)}\bigr)^* \sigma_{2(N'+q),1(N+q)} \ket{2(N')}\bra{1(N)} \Bigr] ,
\end{align}
where $\sigma_{i(N),j(N')}=\bra{i(N)}\sigma\ket{j(N')}$ are the matrix elements of the atom-laser density operator $\sigma$ in the dressed basis. The decay rates appearing in Eq.~\eqref{eq:Ldiss} are determined by the coefficients \eqref{eq:Aq} as $\Gamma_{jj}^{(q)}=\Gamma(q\omega_L) \lvert A_{jj}^{(q)} \rvert^2$ (for $j=1,2$), $\Gamma_{12}^{(q)}=\Gamma(q\omega_L+\Omega) \lvert A_{12}^{(q)} \rvert^2$, and $\Gamma_{21}^{(q)}=\Gamma(q\omega_L-\Omega) \lvert A_{21}^{(q)} \rvert^2$. Due to the properties of the form factor $\Gamma(\omega)$, dissipative $j\to j$ and $2\to 1$ transitions are possible only if $q>0$, while the $1\to 2$ transition can occur also in the case $q=0$, which is particularly interesting since the frequency of the emitted photons can be much smaller than the pump frequency $\omega_L$. As we will show in the following, such a transition can be determined by the presence of permanent dipole moments. The transitions with the lowest $q$ determined by the dissipative dynamics \eqref{eq:Ldiss} are graphically represented in Figure \ref{fig:transitions}. The strength of coherence transfer terms between dressed ``1'' and ``2'' states in Eq.~\eqref{eq:Ldiss} is instead determined by $K_{12}^{(q)} = \sqrt{\Gamma_{11}^{(q)}\Gamma_{22}^{(q)}}\ee^{\ii \varphi(q)}$, where the phase factor depends on the relative phase between $(A_{11}^{(q)})^*$ and $A_{22}^{(q)}$. The detailed derivation of all the term in Eq.~\eqref{eq:Ldiss} is reported in the Appendix.

The approximate invariance of atom-laser couplings with respect to $N$ due to assumption \eqref{eq:laserstat}, expressed in the $N$-independence of the amplitudes $A_{ij}^{(q)}$ and \textit{a fortiori} of all the coefficients appearing in Eq.~\eqref{eq:Ldiss}, allows reducing the dynamics of the atom-laser system, complicated even in the standard RWA case, to a solvable set of differential equations. Specifically, one can define the reduced coherences $\sigma_{ij}^{(\ell)} = \sum_N \sigma_{i(N),j(N+\ell)}$, which, for the case $i\neq j$, follow decoupled evolution equations:
\begin{align}
    \dot{\sigma}_{12}^{(\ell)} & = \left[ \ii (\ell\omega_L - \Omega) - \Gamma_{\mathrm{coh}} \right] \sigma_{12}^{(\ell)} , \\
    \dot{\sigma}_{21}^{(\ell)} & = \left[ \ii (\ell\omega_L + \Omega) - \Gamma_{\mathrm{coh}} \right] \sigma_{12}^{(\ell)} ,
\end{align}
with
\begin{equation}\label{eq:Gammacoh}
    \Gamma_{\mathrm{coh}} = \frac{1}{2} \sum_{q,i,j} \Gamma_{ij}^{(q)} - \mathrm{Re} \left( \sum_{q} K_{12}^{(q)} \right) 
\end{equation}
being the $\ell$-independent decay rate of such terms. Notice that the sum of the imaginary parts of $K_{12}^{(q)}$ provides a correction of $O(\Gamma_0)$ to the imaginary part of the coefficients, which can be consistently neglected due to the condition \eqref{eq:enscales}. Instead, the evolution equations of the equal-index reduced coherences are coupled to each other,
\begin{align}
    \dot{\sigma}_{11}^{(\ell)} & = \left[ \ii \ell\omega_L - \Gamma_{12} \right] \sigma_{11}^{(\ell)} + \Gamma_{21} \sigma_{22}^{(\ell)}  \\
    \dot{\sigma}_{22}^{(\ell)} & = \left[ \ii \ell\omega_L - \Gamma_{21} \right] \sigma_{22}^{(\ell)} + \Gamma_{12} \sigma_{11}^{(\ell)}
\end{align}
with
\begin{equation}\label{eq:Gammaij}
    \Gamma_{ij} = \sum_{q} \Gamma_{ij}^{(q)} .
\end{equation}
As opposed to the case $i\neq j$, all the above equations admit a steady-state solution. In particular, the quantities $\Pi_{1}=\sigma_{11}^{(0)}$ and $\Pi_{2}=\sigma_{22}^{(0)}$, representing the total population of the ``1'' and ``2'' dressed states, respectively, approach the asymptotic values
\begin{equation}\label{eq:Pstat}
    \Pi_1^{\mathrm{st}} = \frac{\Gamma_{21}}{\Gamma_{\mathrm{pop}}}, \quad \Pi_2^{\mathrm{st}} = \frac{\Gamma_{12}}{\Gamma_{\mathrm{pop}}}, \quad \Gamma_{\mathrm{pop}} = \Gamma_{12}+\Gamma_{21},
\end{equation}
determined by the balance between dissipative transitions that change the dressed state label. The quantity $\Gamma_{\mathrm{pop}}$ defined in Eq.~\eqref{eq:Pstat} is not a mere normalization factor, but represents also the rate at which populations approach the steady state. The same-index coherences follow a similar dynamics, asymptotically evolving as $\sigma_{jj}^{(\ell)}(t)=\ee^{\ii\ell \omega_L}\Pi_j^{\mathrm{st}}$. It is worth noticing that the reduced evolution equations appear identical to the standard dressed-atom treatment, despite the population transition rates $\Gamma_{ij}$ and the coherence decay rate $\Gamma_{\mathrm{coh}}$ contain in principle contributions from an infinite number of transitions. The reason of such correspondence essentially lies in hypotheses \eqref{eq:laserstat}-\eqref{eq:enscales} and are independent of the model details.

The asymptotic spectral density of emitted photons can be evaluated using the field operator evolution in the Heisenberg picture, the condition \eqref{eq:enscales} of emission line separation, and the values of steady-state populations: 
\begin{align}\label{eq:spectral}
    \mathcal{J}(\omega) & = \frac{\Gamma(\omega)}{\pi} \nonumber \\ 
    & \times \mathrm{Re} \lim_{T\to\infty} \int_0^{T} \frac{\dd t}{T} \int_0^{\infty} \dd \tau \langle \mathcal{S}_+(t+\tau)\mathcal{S}_-(t) \rangle \ee^{-\ii\omega\tau} \nonumber \\
    & = \sum_{q} \left( \mathcal{J}_{+}^{(q)}(\omega) + \mathcal{J}_{-}^{(q)}(\omega) + \mathcal{J}_{c}^{(q)}(\omega) \right) = \sum_q \mathcal{J}^{(q)}(\omega) .
\end{align}
The details of the derivation are reported in the Appendix. In Eq.~\eqref{eq:spectral}, the average is computed on an initial state in the form $\sigma(0)\otimes\ket{\mathrm{vac}}\bra{\mathrm{vac}}$, where $\sigma(0)$ is the initial state of the atom-laser system and $a(\omega)\ket{\mathrm{vac}}=0$ for all $\omega$'s. Time average is made necessary by the presence of persistent contributions to the correlator in the steady-state regime. At the same time, the limit allows to retain only the contributions that scale with $T$, averaging out the oscillating terms. Notice that the latter issue is not present when the radiative cascade is treated in the RWA case \cite{cohentannoudjiAPI}.

The evaluation of the spectral density \eqref{eq:spectral}, described in detail in the Appendix, makes use of the quantum regression theorem, consistent with the Markovian treatment of the atom-radiation interaction \cite{cohentannoudjiAPI}. The contribution of sideband photons, for any value of $q$, consists in the Lorentzian lines
\begin{align}
    \mathcal{J}_{+}^{(q)}(\omega) & = \frac{\Gamma_{12}^{(q)} \Pi_1^{\mathrm{st}}}{\pi} \frac{\Gamma_{\mathrm{coh}}}{[\omega - (q\omega_L + \Omega) ]^2 + \Gamma_{\mathrm{coh}}^2 } , \\
    \mathcal{J}_{-}^{(q)}(\omega) & = \frac{\Gamma_{21}^{(q)} \Pi_2^{\mathrm{st}}}{\pi} \frac{\Gamma_{\mathrm{coh}}}{[\omega - (q\omega_L - \Omega) ]^2 + \Gamma_{\mathrm{coh}}^2 } ,
\end{align}
all characterized by the same width $\Gamma_{\mathrm{coh}}$, which can be integrated to provide the total line intensities
\begin{equation}\label{eq:sidebands}
    I_+^{(q)} = \Gamma_{12}^{(q)} \Pi_1^{\mathrm{st}} , \qquad I_-^{(q)} = \Gamma_{21}^{(q)} \Pi_2^{\mathrm{st}} .
\end{equation}
The central lines have a more complicated structure,
\begin{eqnarray}
    \mathcal{J}_{c}^{(q)}(\omega) & = & \left[ \Gamma_{11}^{(q)} \left( \Pi_1^{\mathrm{st}} \right)^2 + \Gamma_{22}^{(q)} \left( \Pi_2^{\mathrm{st}} \right)^2 + 2\,\mathrm{Re} K_{12}^{(q)} \Pi_1^{\mathrm{st}} \Pi_2^{\mathrm{st}} \right] \nonumber \\
    & & \times \delta(\omega - q\omega_L) \nonumber \\
    & + & \left( \Gamma_{11}^{(q)} + \Gamma_{22}^{(q)} - 2\,\mathrm{Re} K_{12}^{(q)} \right) \Pi_1^{\mathrm{st}} \Pi_2^{\mathrm{st}} \nonumber \\
    & & \times \frac{1}{\pi} \frac{\Gamma_{\mathrm{pop}}}{( \omega - q\omega_L )^2 + \Gamma_{\mathrm{pop}}^2 } ,
\end{eqnarray}
with the coherent peaks at $\omega=q\omega_L$ surrounded by Lorenzian lines whose width corresponds, for any $q$, to $\Gamma_{\mathrm{pop}}$. It is worth noticing that such a component of the spectral density depends on the coherence transfer coefficients $K_{12}^{(q)}$ only through their real parts, in accordance with the independently obtained result in Eq.~\eqref{eq:Gammacoh}. Integration over a central line erases the dependence on $\mathrm{Re} K_{12}^{(q)}$ (hence, of the relative phase between $A_{11}^{(q)}$ and $A_{22}^{(q)}$), leading to the intuitive form
\begin{equation}
    I_c^{(q)} = \Gamma_{11}^{(q)} \Pi_1^{\mathrm{st}} + \Gamma_{22}^{(q)} \Pi_2^{\mathrm{st}} .
\end{equation}
These results open the possibility to compare the strength of emission lines for any specific atom-laser Hamiltonian satisfying the phenomenological assumptions. Interestingly, the ratios of transition line weights (as well as the steady state populations) do not depend on the magnitude of $\Gamma_0$, which can be considered arbitrarily small to satisfy condition \eqref{eq:enscales}. In particular, one can compare the strength of the small-freqency singlet around $\omega=\Omega$ with that of the higher-energy triplets, obtaining 
\begin{equation}
    \frac{I_+^{(0)}}{I^{(q)}} = \frac{\Gamma_{12}^{(0)} \Gamma_{21}}{ \left( \Gamma_{11}^{(q)} + \Gamma_{12}^{(q)} \right) \Gamma_{21} + \left( \Gamma_{22}^{(q)} + \Gamma_{21}^{(q)} \right) \Gamma_{12} } ,
\end{equation}
with $I^{(q)}=I^{(q)}_c + I^{(q)}_+ + I^{(q)}_+$ the total triplet strength. The above derivation of the spectral characteristics of the radiative emission from the dressed system is performed with a minimal set of assumptions \eqref{eq:laserstat}--\eqref{eq:enscales}, and is Hamiltonian-independent: The Hamiltonian enters the description at the stage of evaluation of the model parameters $A_{ij}^{(q)}$ according to Eq.~\eqref{eq:Aq}.

\begin{table}
    \centering
    \begin{tabular}{cccc}
         \toprule
         \textbf{Multiplet}\, & \textbf{Number}\, & \textbf{Central}\, & \textbf{Line weight}\, \\
         \textbf{label}\, & \textbf{of lines}\, & \textbf{frequency}\, & \textbf{scaling}\, \\
         \midrule
         $q=0$ & $1$ & $\Omega$ & $ \left(\frac{\Omega_{\mathrm{AS}}}{\omega_L} \right)^2 \left( \frac{\Omega}{\omega_L} \right)^{\alpha}$ \\ 
         $q=1$ & $3$ & $\omega_L$ & $1$ \\ 
         $q=2$ & $3$ & $2\omega_L$ & $ \left(\frac{\Omega_{\mathrm{AS}}}{\omega_L} \right)^2$ \\ 
         $q=3$ & $3$ & $3\omega_L$ & $ \left(\frac{\Omega_{\mathrm{R}}}{\omega_L} \right)^2 \left( \frac{\Omega}{\omega_L} \right)^{2}$ \\ 
         \bottomrule
    \end{tabular}
    \caption{Properties of lowest-energy emission multiplets in steady-state radiative cascade. RWA atom-laser interaction induces doublet splitting $\Omega$. Perturbation due to atomic permanent dipole moment and counter-rotating terms gives rise to new lines with magnitudes scaled with $\Omega_{\mathrm{AS}}$ and $\Omega_{\mathrm{R}}$, respectively.}
    \label{tab:lines}
\end{table}

\begin{figure}
    \centering
    \includegraphics[width=0.49\textwidth]{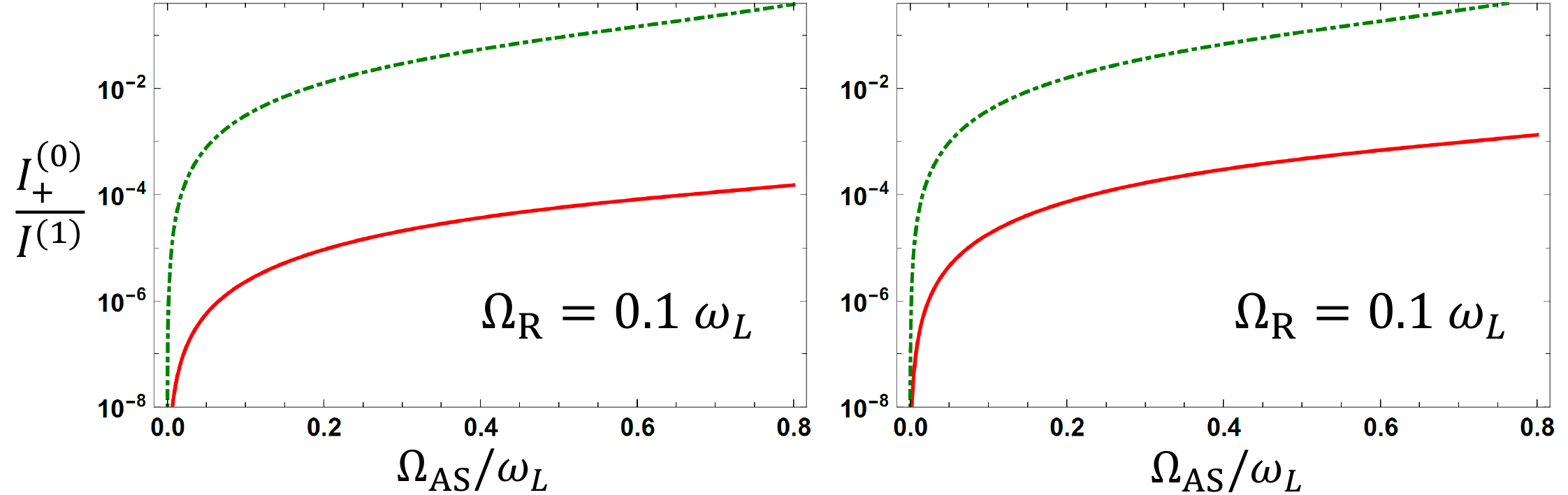}
    \caption{Relative weight of the low-frequency emission ($q=0$) compared to the Mollow triplet ($q=1$), computed for the atom-laser Hamiltonian \eqref{eq:HAL} with $\omega_0=\omega_L$, and for $\Gamma(\omega)\propto\omega^3$. The numerical evaluation based on Eq.~\eqref{eq:ratio01} (solid red lines) is compared with the result obtained neglecting the dependence of decay rates on the dressed state energies (dot-dashed green lines). The emission intensity for all cases tends to 0 as $\Omega_\mathrm{AS}\rightarrow 0$.}
    \label{fig:intensity}
\end{figure}

After deriving the general results, let us focus on the case of an atom-laser Hamiltonian
\begin{align}\label{eq:HAL}
    H_{\mathrm{AL}} & = \omega_0 \ket{e}\bra{e}\otimes \openone_L + \openone_A\otimes \sum_N N\omega_L \ket{N}\bra{N} \nonumber \\
    & + \left[ \frac{\Omega_{\mathrm{R}}}{2} \left( \ket{e}\bra{g} + \ket{g}\bra{e} \right) + \Omega_{\mathrm{AS}} \ket{e}\bra{e} \right] \otimes h_L ,  
\end{align}
where $h_L=\sum_N \left( \ket{N}\bra{N-1} + \ket{N-1}\bra{N} \right)$, and $\openone_L$ ($\openone_A$) is the identity operator on the laser (atom). The Rabi frequency $\Omega_{\mathrm{R}}$ and the coefficient $\Omega_{\mathrm{AS}}$, describe respectively the laser coupling with the transition dipole moments proportional to $\ket{e}\bra{g},\,\ket{g}\bra{e}$ and the permanent dipole moments of the excited state $\ket{e}\bra{e}$. They are both proportional to $\sqrt{\langle N \rangle}$. Let us consider the case of $\Omega_{\mathrm{AS}},\Omega_{\mathrm{R}}\ll\omega_L$, when permanent-dipole and counter-rotating terms can be viewed as a correction to RWA. Notice that these conditions are unrelated to the request \eqref{eq:enscales}, since $\Omega$ is mostly determined by the RWA terms. Considering the composition of the dressed states, we obtain the lowest-order relative weight of the low-frequency singlet with respect to the main triplet as
\begin{equation}\label{eq:ratio01}
    \frac{I_+^{(0)}}{I^{(1)}} \simeq \frac{\Gamma(\Omega)}{\Gamma(\omega_L)} \left( \frac{\Omega_{\mathrm{AS}}}{\omega_L} \right)^2 \frac{\Omega_\mathrm{R}^2}{4\left[ \Omega_{\mathrm{R}}^2 + (\omega_0 - \omega_L)^2 \right] } .
\end{equation}
As shown in the Appendix, such an expression remains effective with good approximation in a range that goes beyond the $\Omega_{\mathrm{R}},\Omega_{\mathrm{AS}}\ll\omega_L$ regime. The low-frequency transition is naturally suppressed by a factor of $(\Omega_{\mathrm{AS}}/\omega_L)^2$, due to the dependence of $\Gamma_{12}^{(0)}$ on the permanent-dipole perturbation. Moreover, suppose one desires emission at a frequency $\Omega\ll\omega_L$. In that case, the transition is also at risk of heavy suppression by the behavior of the interaction form factor, which, in homogeneous media and in the range of validity of the dipole approximation, scales like $\omega^3$. For the same reason, the low-energy singlet competes even with the $q=2$ triplet, which has the same scaling with $\Omega_{\mathrm{AS}}$:
\begin{equation}
    \frac{I_+^{(0)}}{I^{(2)}} \simeq \frac{\Gamma(\Omega)}{\Gamma(2\omega_L)} \frac{\Omega_\mathrm{R}^2}{4 \left[ \Omega_{\mathrm{R}}^2 + (\omega_0 - \omega_L)^2 \right]} .
\end{equation}
Though affected by an even amplified problem with the form factor, comparison with the $q=3$ triplet is mitigated by the fact that the weight of the latter is proportional to the fourth power of $\Omega/\omega_L$ (see Appendix), yielding
\begin{equation}
    \frac{I_+^{(0)}}{I^{(3)}} \sim \frac{\Gamma(\Omega)}{\Gamma(3\omega_L)} \left( \frac{\Omega_{\mathrm{AS}}}{\omega_L} \right)^2 \left( \frac{\omega_L}{\Omega} \right)^{4} \sim \left(\frac{\Omega_{\mathrm{AS}}}{\Omega} \right)^2 \left( \frac{\Omega}{\omega_L} \right)^{\alpha-2}
\end{equation}
in the case of a form factor behaving like $\Gamma(\omega)\sim\omega^{\alpha}$. The $\alpha$ parameter can be tailored, e.g., by exploiting low-dimensional geometries. The parameters characterizing the lowest-frequency emission lines in the perturbative case are reported in Table~\ref{tab:lines}. Enhancing the asymptotic production of low-frequency photons is therefore a difficult task in free space, but can be obtained if the form factor is properly tailored to be peaked at energies around $\hbar\Omega$. One possibility to perform such an enhancement would be to buffer the coupling between the atom and radiation propagating in free space with a lossy resonant cavity (see also discussions in Ref.~\cite{scala2021beyond}). 
Inspection of the spectral contributions provides a further quality factor of the emission lines, highlighting their prominence above the background \cite{scala2021beyond}. Focusing again on the low-frequency transition, we can compare the peak value $\mathcal{J}_+^{(0)}(\Omega)$ with the total tail contributions generated by the remaining transition at the same frequency. Assuming that the $q=1$ triplet is dominant, and neglecting the difference among its frequencies, the quality factor reads
\begin{equation}
    \frac{\mathcal{J}_+^{(0)}(\Omega)}{ \sum_{q>0} \mathcal{J}^{(q)}(\Omega) } \simeq \left( \frac{\omega_L}{\Gamma_{\mathrm{coh}}} \right)^2 \frac{ \Gamma_{12}^{(0)} }{ \Gamma_{12}^{(1)} } \frac{1}{ 4 - \Gamma_{\mathrm{pop}}/\Gamma_{\mathrm{coh}} } ,
\end{equation}
highlighting the fact that it can be made larger by decreasing the dipole transition coupling constant, which is not fixed by the parameters entering the relative integrated peak intensities. From all the above results, the crucial role of dressed-state decay rates is evident. When transitions occur on different energy scales, neglecting their dependence on the coupling form factor which comes from knowledge of the dressed states can lead to orders-of-magnitude errors in the evaluation of photon production rates, as shown in Figure~\ref{fig:intensity}.

The framework introduced in this Letter extend the dressed-atom analysis of a steady-state radiative cascade to the case of an arbitrary coupling between atom and laser. The conditions for the validity of the results here presented are not related to the specific form of the interaction Hamiltonian, but to spectral features and relations between energy scales. Given these properties, we obtain a general form of the spectral density of all the emerging fluorescence multiplets, from which the emission intensity at different frequencies can be determined. At the same time, it is worth noticing that the fully quantum treatment of the atom-laser system provides information that does not emerge if the laser mode is treated as a classical field, such as the dependence of line weights on the on-shell electromagnetic form factors and on the atom-laser eigenstate mixing, which determines the coefficients in Eq.~\eqref{eq:Aq} and the role of interference effects in determining the coherence decay rates. Perspectives in this research line can involve the generalization to many-emitter ensembles, thus extending the semiclassical treatment in Refs.~\cite{chestnov2017,gladysz2020}, and the introduction of different dissipation mechanisms or interaction with more than one laser mode \cite{paspalakis2013,anton2017,gladysz2020}. A qualitatively interesting extension of this work would be to investigate the case of a multi-level atom, where more low-energy lines can emerge, possibly with new coherence effects in the case in which more than two or more transitions inside a multiplet occur at the same frequency.

\begin{acknowledgments}
\textit{Acknowledgments.---} F.V.P. acknowledges support by the PNRR MUR project PE0000023-NQSTI and by the INFN project QUANTUM. K.S. is grateful for the support by the National Science Centre, Poland (Project No.\ 2020/39/I/ST3/00526).
\end{acknowledgments}

\appendix

\onecolumngrid

\section{Derivation of the master equation}

The dissipative dynamics of the dressed atom is determined by the atom-radiation interaction Hamiltonian
\begin{equation}\label{eq:VAR}
    V_{\mathrm{AR}} = \hbar \int \dd\omega \, \sqrt{\frac{\Gamma(\omega)}{2\pi}} \left ( \mathcal{S}_+ a(\omega) + \mathcal{S}_- a^{\dagger}(\omega) \right) , 
\end{equation}
where $\Gamma(\omega)$ vanishes if $\omega\leq 0$. Due to the structure of the atom-laser Hamiltonian
\begin{equation}\label{eq:HAL}
    H_{\mathrm{AL}} = \sum_{j=1,2}\sum_N E_j(N) \ket{j(N)}\bra{j(N)} ,  \quad \text{with } E_j(N) = \hbar \left( N\omega_L - (-1)^{j}\frac{\Omega}{2} \right) ,
\end{equation}
the atom-radiation dynamics can yield on-shell decay of the dressed eigenstates at the frequencies $q\omega_L$ and $q\omega_L\pm\Omega$. 

Using the assumption that, for all practical purposes, the composition of the dressed eigenstates $\ket{j(N)}$ in terms of the bare states $\ket{e,N'}$ and $\ket{g,N'}$ only depends on $N-N'$, the atomic transition operators $\mathcal{S}_{+}=\ket{e}\bra{g}$ and $\mathcal{S}_{-}=\ket{g}\bra{e}$ can be conveniently expressed as
\begin{equation}\label{eq:Spm}
    \mathcal{S}_{+} = \sum_{i,j\in\{1,2\}} \sum_q A_{ij}^{(q)} S_+^{ij(q)}, \qquad \mathcal{S}_{-} = \sum_{i,j\in\{1,2\}} \sum_q \bigl(A_{ij}^{(q)}\bigr)^* S_-^{ij(q)}
\end{equation}
with the coefficients
\begin{equation}\label{eq:Aq}
    A_{ij}^{(q)} = \sum_p \langle i(N+p+q)|e,N \rangle \langle g,N|j(N+p) \rangle 
\end{equation}
actually independent of $N$, and
\begin{equation}
    S_+^{ij(q)} = \sum_N \ket{i(N)}\bra{j(N-q)}, \qquad S_-^{ij(q)} = \sum_N \ket{j(N-q)}\bra{i(N)} .
\end{equation}
The Lindblad operators associated to the allowed transitions can thus be expressed as
\begin{align}
    L_{+}^{(q)} & = \sum_N \sqrt{ \Gamma \left( \frac{E_1(N)-E_2(N-q)}{\hbar} \right) } \bigl(A_{12}^{(q)}\bigr)^* \ket{2(N-q)}\bra{1(N)} = \sqrt{\Gamma(q\omega_L+\Omega)} \bigl(A_{12}^{(q)}\bigr)^* S_-^{12(q)} , \\
    L_{-}^{(q)} & = \sum_N \sqrt{ \Gamma \left( \frac{E_2(N)-E_1(N-q)}{\hbar} \right) } \bigl(A_{21}^{(q)}\bigr)^* \ket{1(N-q)}\bra{2(N)} = \sqrt{\Gamma(q\omega_L-\Omega)} \bigl(A_{21}^{(q)}\bigr)^* S_-^{21(q)} , \\
    L_{c}^{(q)} & = \sum_{jN} \sqrt{ \Gamma \left( \frac{E_j(N)-E_j(N-q)}{\hbar} \right) } \bigl(A_{jj}^{(q)}\bigr)^* \ket{j(N-q)}\bra{j(N)} = \sqrt{\Gamma(q\omega_L)} \left[ \bigl(A_{11}^{(q)}\bigr)^* S_-^{11(q)} + \bigl(A_{22}^{(q)}\bigr)^* S_-^{22(q)} \right] .
\end{align}
Therefore, the dissipative part of the GKLS master equation, applied to the atom-laser density matrix $\sigma$, reads
\begin{align}\label{eq:Ldiss0}
    \mathcal{L}_{\mathrm{diss}}(\sigma) = & \sum_{n\in\{+,-,c\}} \sum_q \left[ L_n^{(q)}\sigma \bigl(L_n^{(q)}\bigr)^{\dagger} - \frac{1}{2} \left\{ \bigl(L_n^{(q)}\bigr)^{\dagger} L_n^{(q)}, \sigma \right\} \right] \nonumber \\
    = & \sum_q \left[ \sum_{i,j\in\{1,2\}} \Gamma_{ij}^{(q)} \left( S_-^{ij(q)}\sigma S_+^{ij(q)} - \frac{1}{2} \left\{ S_+^{ij(q)} S_-^{ij(q)}, \sigma \right\} \right) + K_{12}^{(q)} S_-^{11(q)}\sigma S_+^{22(q)} + \bigl( K_{12}^{(q)} \bigr)^* S_-^{22(q)}\sigma S_+^{11(q)} \right] ,
\end{align}
where
\begin{equation}
    \Gamma_{11}^{(q)} = \Gamma(q\omega_L) \bigl\lvert A_{11}^{(q)} \bigr\rvert^2 , \quad \Gamma_{22}^{(q)} = \Gamma(q\omega_L) \bigl\lvert A_{22}^{(q)} \bigr\rvert^2 , \quad \Gamma_{12}^{(q)} = \Gamma(q\omega_L+\Omega) \bigl\lvert A_{12}^{(q)} \bigr\rvert^2 , \quad \Gamma_{21}^{(q)} = \Gamma(q\omega_L-\Omega) \bigl\lvert A_{21}^{(q)} \bigr\rvert^2 ,
\end{equation}
and
\begin{equation}\label{eq:K12}
    K_{12}^{(q)} = \Gamma(q\omega_L) \bigl\lvert A_{11}^{(q)} A_{22}^{(q)} \bigr\rvert \ee^{\ii \varphi(q)} = \sqrt{ \Gamma_{11}^{(q)} \Gamma_{22}^{(q)} } \ee^{\ii \varphi(q)} ,
\end{equation}
with $\varphi(q)$ the relative phase between $A_{22}^{(q)}$ and $\bigl( A_{11}^{(q)} \bigr)^*$ and $\{A,B\}=AB+BA$ the anticommutator in the atom-laser Hilbert space. Expanding the expression \eqref{eq:Ldiss0} in terms of the transition operators $S_{\pm}^{(q)}$ provides
\begin{align}\label{eq:Ldiss1}
    \mathcal{L}_{\mathrm{diss}}(\sigma) = & \sum_{qNN'} \Biggl[ \sum_{ij} \Gamma_{ij}^{(q)} \left( \sigma_{i(N),i(N')} \ket{j(N-q)}\bra{j(N'-q)} - \frac{\delta_{NN'}}{2} \left\{ \ket{i(N)}\bra{i(N)} , \sigma \right\} \right) \nonumber \\
    & + K_{12}^{(q)} \sigma_{1(N),2(N')} \ket{1(N-q)}\bra{2(N'-q)} + \bigl( K_{12}^{(q)} \bigr)^* \sigma_{2(N),1(N')} \ket{2(N-q)}\bra{1(N'-q)} \Biggr] ,
\end{align}
with
\begin{equation}
    \sigma_{i(N),j(N')} = \bra{i(N)} \sigma \ket{j(N')} .
\end{equation}

\section{Full and reduced atom-laser dynamics}

The evolution of the atom-laser density matrix $\sigma$ is determined by the combination of the Markovian dissipative dynamics \eqref{eq:Ldiss1} and the Hamiltonian dynamics yielded by $H_{\mathrm{AL}}$:
\begin{equation}
    \frac{\dd}{\dd t} \sigma(t) = - \frac{\ii}\hbar \left[H_{\mathrm{AL}},\sigma (t) \right] + \mathcal{L}_{\mathrm{diss}}(\sigma(t)) = \frac{\ii}\hbar \sum_{j,N} E_j(N) \left[\sigma (t), \ket{j(N)}\bra{j(N)} \right] + \mathcal{L}_{\mathrm{diss}}(\sigma(t)) .
\end{equation}
From the above expression and from \eqref{eq:Ldiss1}, one can obtain the coupled evolution equations for each element:
\begin{align}
    \dot{\sigma}_{1(N),1(N')} & = \ii(N'-N)\omega_L \sigma_{1(N),1(N')} + \sum_q \left[ \Gamma_{11}^{(q)} \sigma_{1(N+q),1(N'+q)} + \Gamma_{21}^{(q)} \sigma_{2(N+q),2(N'+q)} - \left( \Gamma_{11}^{(q)} + \Gamma_{12}^{(q)} \right) \sigma_{1(N),1(N')} \right] , \\
    \dot{\sigma}_{2(N),2(N')} & = \ii(N'-N)\omega_L \sigma_{2(N),2(N')} + \sum_q \left[ \Gamma_{12}^{(q)} \sigma_{1(N+q),1(N'+q)} + \Gamma_{22}^{(q)} \sigma_{2(N+q),2(N'+q)} - \left( \Gamma_{21}^{(q)} + \Gamma_{22}^{(q)} \right) \sigma_{2(N),2(N')} \right] , \\
    \dot{\sigma}_{1(N),2(N')} & = \ii[(N-N')\omega_L - \Omega] \sigma_{1(N),2(N')} + \sum_q \left[ - \frac{\Gamma_{11}^{(q)} + \Gamma_{12}^{(q)} + \Gamma_{21}^{(q)} + \Gamma_{22}^{(q)}}{2} \sigma_{1(N),2(N')} + K_{12}^{(q)} \sigma_{1(N+q),2(N'+q)} \right] , \\
    \dot{\sigma}_{2(N),1(N')} & = \ii[(N-N')\omega_L + \Omega] \sigma_{2(N),1(N')} + \sum_q \left[ - \frac{\Gamma_{11}^{(q)} + \Gamma_{12}^{(q)} + \Gamma_{21}^{(q)} + \Gamma_{22}^{(q)}}{2} \sigma_{2(N),1(N')} + \bigl(K_{12}^{(q)}\bigr)^* \sigma_{2(N+q),1(N'+q)} \right] .
\end{align}
After introducing the reduced density matrix elements
\begin{equation}\label{eq:sigmared}
    \sigma_{ij}^{(\ell)} = \sum_N \sigma_{i(N),j(N+\ell)} ,
\end{equation}
one can derive their dynamics by summing the above equations on $N$. For the elements with different indices, one gets the decoupled equations
\begin{equation}
    \dot{\sigma}_{12}^{(\ell)} = \left[ \ii (\ell\omega_L - \Omega) - \Gamma_{\mathrm{coh}} \right] \sigma_{12}^{(\ell)}, \qquad \dot{\sigma}_{21}^{(\ell)} = \left[ \ii (\ell\omega_L + \Omega) - \Gamma_{\mathrm{coh}} \right] \sigma_{21}^{(\ell)} 
\end{equation}
with 
\begin{equation}
    \Gamma_{\mathrm{coh}} = \sum_q \left[ \frac{\Gamma_{11}^{(q)} + \Gamma_{12}^{(q)} + \Gamma_{21}^{(q)} + \Gamma_{22}^{(q)}}{2} - \mathrm{Re}\left( K_{12}^{(q)} \right) \right] ,
\end{equation}
which can be integrated as
\begin{equation}\label{eq:sigma12evo}
    \sigma_{12}^{(\ell)}(t+\tau) = \exp\left[ \ii (\ell\omega_L - \Omega)\tau - \Gamma_{\mathrm{coh}}\tau \right] \sigma_{12}^{(\ell)}(t), \quad \sigma_{21}^{(\ell)}(t+\tau) = \exp\left[ \ii (\ell\omega_L + \Omega)\tau - \Gamma_{\mathrm{coh}}\tau \right] \sigma_{21}^{(\ell)}(t),
\end{equation}
respectively. Notice that the imaginary parts of $K_{12}^{(q)}$ have been consistently neglected with respect to the $O(\ell\omega_L\pm\Omega)$ terms. It is also interesting to observe that, in the standard RWA evaluation, the real part of $K_{12}^{(1)}$ is negative, thus providing an increase in the coherence decay rate. The reason will become clear in the last section.

The equal-index reduced density matrix elements satisfy the coupled equations 
\begin{equation}\label{eq:cohjj}
    \dot{\sigma}_{11}^{(\ell)} = \left[ \ii \ell\omega_L - \Gamma_{12} \right] \sigma_{11}^{(\ell)} + \Gamma_{21} \sigma_{22}^{(\ell)} , \qquad
    \dot{\sigma}_{22}^{(\ell)} = \left[ \ii \ell\omega_L - \Gamma_{21} \right] \sigma_{22}^{(\ell)} + \Gamma_{12} \sigma_{11}^{(\ell)} .
\end{equation}
A specific case is represented by the populations 
\begin{equation}
    \Pi_j = \sigma_{jj}^{(0)} = \sum_N \bra{j(N)} \sigma \ket{j(N)} \qquad \text{with } j=1,2,
\end{equation}
whose evolution, satisfying $\Pi_1(t)+\Pi_2(t)=1$ at all times, converges to the steady-state values
\begin{equation}\label{eq:Pstat}
    \Pi_1^{\mathrm{st}} = \frac{\Gamma_{21}}{\Gamma_{\mathrm{pop}}}, \qquad \Pi_2^{\mathrm{st}} = \frac{\Gamma_{12}}{\Gamma_{\mathrm{pop}}}, \qquad \text{with }\Gamma_{\mathrm{pop}} = \Gamma_{12}+\Gamma_{21} ,
\end{equation}
while the coherences approach the steady-state behavior
\begin{equation}
    \sigma_{jj,\mathrm{st}}^{(\ell)} (t) = \ee^{ \ii \ell\omega_L t } \Pi_{j}^{\mathrm{st}} .
\end{equation}
In general, the coupled equations \eqref{eq:cohjj} can be integrated into
\begin{align}
    \sigma_{11}^{(\ell)}(t+\tau) & = \ee^{ \ii \ell\omega_L \tau } \left[ \left( \Pi_1^{\mathrm{st}} + \Pi_2^{\mathrm{st}} \ee^{-\Gamma_{\mathrm{pop}}\tau} \right) \sigma_{11}^{(\ell)}(t) + \Pi_1^{\mathrm{st}} \left( 1 - \ee^{-\Gamma_{\mathrm{pop}}\tau} \right) \sigma_{22}^{(\ell)}(t) \right] , \label{eq:coh11} \\
    \sigma_{22}^{(\ell)}(t+\tau) & = \ee^{ \ii \ell\omega_L \tau } \left[ \Pi_2^{\mathrm{st}} \left( 1 - \ee^{-\Gamma_{\mathrm{pop}}\tau} \right) \sigma_{11}^{(\ell)}(t) + \left( \Pi_2^{\mathrm{st}} + \Pi_1^{\mathrm{st}} \ee^{-\Gamma_{\mathrm{pop}}\tau} \right) \sigma_{22}^{(\ell)}(t) \right] . \label{eq:coh22}
\end{align}
Therefore, the quantity $\Gamma_{\mathrm{pop}}$ defined in Eq.~\eqref{eq:Pstat} plays the role of a relaxation rate towards the steady-state regime.

\section{Spectral density of emitted photons}

The spectral density of the emitted photons can be obtained starting from the Heisenberg evolution of the field operators \cite{cohentannoudjiAPI} entailed by the Hamiltonian
\begin{equation}
    H = H_{\mathrm{AL}} + \hbar \int \dd\omega\, \omega a^{\dagger}(\omega) a(\omega) + \hbar \int \dd\omega \, \sqrt{\frac{\Gamma(\omega)}{2\pi}} \left ( \mathcal{S}_+ a(\omega) + \mathcal{S}_- a^{\dagger}(\omega) \right) ,
\end{equation}
yielding
\begin{equation}
    a(\omega,t) = a(\omega) \ee^{-\ii\omega t} - \ii \sqrt{\frac{\Gamma(\omega)}{2\pi}} \int_0^t \dd t'\, \mathcal{S}_-(t') \ee^{-\ii\omega(t-t')} ,
\end{equation}
where the integration constants are set in such a way that at the initial time $t=0$, when the state of the atom-laser-radiation system is
\begin{equation}
    \rho(0) = \sigma(0) \otimes \ket{\mathrm{vac}}\bra{\mathrm{vac}}
\end{equation}
with $a(\omega)\ket{\mathrm{vac}}=0$, the Heisenberg operator $a(\omega,0)$ coincides with the Schr\"odinger operator $a(\omega)$, acting only on the radiation degrees of freedom.

After an evolution time $T$, the probability density associated to photon frequency reads
\begin{align}
    \mathrm{Tr}\left[ a^{\dagger}(\omega)a(\omega)\rho(T) \right] = \langle a^{\dagger}(\omega,T)a(\omega,T) \rangle & = \frac{\Gamma(\omega)}{2\pi} \int_0^T \dd t \int_0^T \dd t' \langle \mathcal{S}_+(t')\mathcal{S}_-(t) \rangle \ee^{-\ii\omega(t'-t)} \nonumber \\
    & = \frac{\Gamma(\omega)}{\pi} \mathrm{Re} \int_0^T \dd t \int_0^{T-t} \dd \tau \langle \mathcal{S}_+(t+\tau)\mathcal{S}_-(t) \rangle \ee^{-\ii\omega\tau}
\end{align}
with $\langle\dots\rangle$ the average on the initial state $\rho(0)$. The evolution towards a steady state entails the presence of constant contributions in $t$, yielding diverging integrals, and oscillating ones, which are not integrable but provide comparatively negligible contributions. Therefore, it is convenient to define the asymptotic spectral density as
\begin{equation}\label{eq:spectrallim}
    \mathcal{J}(\omega) = \lim_{T\to\infty} \frac{\langle a^{\dagger}(\omega,T)a(\omega,T) \rangle}{T} = \lim_{T\to\infty} \frac{\Gamma(\omega)}{\pi} \mathrm{Re} \int_0^T \frac{\dd t}{T} \int_0^{T-t} \dd \tau \langle \mathcal{S}_+(t+\tau)\mathcal{S}_-(t) \rangle \ee^{-\ii\omega\tau} .
\end{equation}
The correlator in the integral can be evaluated by using the quantum regression theorem (QRT) \cite{cohentannoudjiAPI}, which is consistent with the applicability condition of the Markovian approximation, namely the fact that the time scale of vacuum correlations (in this case, the width of the Fourier transform of $\Gamma(\omega)$) is much smaller than any other scale of time variation in the open system dynamics. Since $\mathcal{S}_+$ can be written as
\begin{equation}\label{eq:Sijq}
    \mathcal{S}_+ = \sum_{i,j\in\{1,2\}} \sum_q \mathcal{S}_+^{ij(q)}, \qquad \text{with } \mathcal{S}_+^{ij(q)} = A_{ij}^{(q)} \sum_{N} \ket{i(N)}\bra{j(N-q)} ,
\end{equation}
the spectral density can be accordingly decomposed in the following terms,
\begin{equation}
    \mathcal{J}(\omega) = \sum_{q} \left( \mathcal{J}_{+}^{(q)}(\omega) + \mathcal{J}_{-}^{(q)}(\omega) + \mathcal{J}_{c}^{(q)}(\omega) \right)
\end{equation}
which will be now derived separately.

\subsection{Sidebands}

From the definitions \eqref{eq:Sijq} and \eqref{eq:sigmared}, one can easily obtain that
\begin{equation}
    \langle \mathcal{S}_+^{ij(q)} \rangle = A_{ij}^{(q)} \sigma_{ji}^{(q)} .
\end{equation}
Therefore, the evolution \eqref{eq:sigma12evo} of the unequal-index coherences determines
\begin{equation}
    \langle \mathcal{S}_+^{12(q)} (t+\tau) \rangle = \exp\left[ \ii (q\omega_L + \Omega)\tau - \Gamma_{\mathrm{coh}}\tau \right] \langle \mathcal{S}_+^{12(q)} (t) \rangle .
\end{equation}
From QRT, $\langle \mathcal{S}_+^{ij(q)} (t+\tau) \mathcal{S}_-(t) \rangle$ follows the same evolution in $\tau$ as $\langle \mathcal{S}_+^{ij(q)} (t+\tau) \rangle$, and therefore
\begin{equation}
    \langle \mathcal{S}_+^{12(q)} (t+\tau) \mathcal{S}_-(t) \rangle = \exp\left[ \ii (q\omega_L + \Omega)\tau - \Gamma_{\mathrm{coh}}\tau \right] \langle \mathcal{S}_+^{12(q)} (t) \mathcal{S}_-(t) \rangle .
\end{equation}
The expectation value of $\mathcal{S}_+^{12(q)} \mathcal{S}_-$ at time $t$ can be evaluated by using the expression \eqref{eq:Spm}, as
\begin{equation}
    \langle \mathcal{S}_+^{12(q)} (t) \mathcal{S}_-(t) \rangle = A_{12}^{(q)} \sum_{j\in\{1,2\}} \sum_{\ell} \bigl( A_{j2}^{(\ell)} \bigr)^* \sigma_{j1}^{(q-\ell)} (t) = A_{12}^{(q)} \sum_{\ell} \bigl( A_{12}^{(\ell)} \bigr)^* \Pi_1^{\mathrm{st}} \ee^{\ii(q-\ell)\omega_L t} + f_{12}^{(q)} (t) ,
\end{equation}
where $f_{12}^{(q)} (t)$ contains transient contributions in $t$, which certainly do not contribute to the spectral density, as defined in \eqref{eq:spectrallim}. Combining the above results, one obtains the spectral density contribution
\begin{align}
    \mathcal{J}_{+}^{(q)}(\omega) & = \lim_{T\to\infty} \frac{\Gamma(\omega)}{\pi} \mathrm{Re} \int_0^T \dd t \int_0^{T-t} \dd \tau \langle \mathcal{S}_+^{12(q)}(t+\tau)\mathcal{S}_-(t) \rangle \ee^{-\ii\omega\tau} \nonumber \\
    & = \frac{\Gamma(\omega)}{\pi} \bigl\lvert A_{12}^{(q)} \bigr\rvert^2 \Pi_1^{\mathrm{st}} \, \mathrm{Re} \int_0^{\infty} \dd \tau \, \ee^{-\ii(\omega-q\omega_L - \Omega) \tau - \Gamma_{\mathrm{coh}}\tau} \nonumber \\
    & = \Gamma_{12}^{(q)} \Pi_1^{\mathrm{st}} \frac{1}{\pi} \frac{\Gamma_{\mathrm{coh}}}{ [\omega - (q\omega_L+\Omega)]^2 + \Gamma_{\mathrm{coh}}^2}
\end{align}
with the last equality holding up to $O(\Gamma_{\mathrm{coh}})$ corrections, considering that $\Gamma_{12}^{(q)}=\Gamma(q\omega_L + \Omega) \bigl\lvert A_{12}^{(q)}\bigr\rvert^2$. The computation of the lower-energy sideband contribution
\begin{equation}
    \mathcal{J}_{-}^{(q)}(\omega) = \lim_{T\to\infty} \frac{\Gamma(\omega)}{\pi} \mathrm{Re} \int_0^T \dd t \int_0^{T-t} \dd \tau \langle \mathcal{S}_+^{21(q)}(t+\tau)\mathcal{S}_-(t) \rangle \ee^{-\ii\omega\tau} = \Gamma_{21}^{(q)} \Pi_2^{\mathrm{st}} \frac{1}{\pi} \frac{\Gamma_{\mathrm{coh}}}{ [\omega - (q\omega_L-\Omega)]^2 + \Gamma_{\mathrm{coh}}^2}
\end{equation}
proceeds in an entirely analogous way, with the roles of indexes 1 and 2 exchanged, and the sign in front of $\Omega$ reversed.

\subsection{Central lines}

The evaluation of the central-line contributions
\begin{equation}
    \mathcal{J}_{c}^{(q)}(\omega) = \lim_{T\to\infty} \frac{\Gamma(\omega)}{\pi} \mathrm{Re} \int_0^T \dd t \int_0^{T-t} \dd \tau \langle \left[\mathcal{S}_+^{11(q)}(t+\tau) + \mathcal{S}_+^{22(q)}(t+\tau)\right]\mathcal{S}_-(t) \rangle \ee^{-\ii\omega\tau}
\end{equation}
is complicated by the fact that the evolutions of $\langle \mathcal{S}_+^{jj(q)} \rangle = A_{jj}^{(q)} \sigma_{jj}^{(q)}$ for $j=1$ and $2$ are coupled to each other. Actually, following the evolution \eqref{eq:coh11}-\eqref{eq:coh22} of the same-index reduced coherences, one obtains 
\begin{equation}
    \langle \mathcal{S}_+^{11(q)}(t+\tau) + \mathcal{S}_+^{22(q)}(t+\tau) \rangle = \ee^{\ii q \omega_L \tau} \left[ \left( C_1^{(q)} + D_1^{(q)} \ee^{-\Gamma_{\mathrm{pop}}\tau} \right) \langle \mathcal{S}_+^{11(q)}(t) \rangle + \left( C_2^{(q)} + D_2^{(q)} \ee^{-\Gamma_{\mathrm{pop}}\tau} \right) \langle \mathcal{S}_+^{22(q)}(t) \rangle \right]
\end{equation}
with
\begin{equation}
    C_1^{(q)} = \Pi_1^{\mathrm{st}} + \frac{A_{22}^{(q)}}{A_{11}^{(q)}} \Pi_2^{\mathrm{st}}, \quad C_2^{(q)} = \Pi_2^{\mathrm{st}} + \frac{A_{11}^{(q)}}{A_{22}^{(q)}} \Pi_1^{\mathrm{st}}, \quad D_1^{(q)} = \Pi_1^{\mathrm{st}} \left( 1 - \frac{A_{22}^{(q)}}{A_{11}^{(q)}} \right), \quad D_2^{(q)} = \Pi_2^{\mathrm{st}} \left( 1 - \frac{A_{11}^{(q)}}{A_{22}^{(q)}} \right) .
\end{equation}
Notice that in the standard dressed-atom approach, in which RWA is considered in the atom-laser interaction, such a computation is oversimplified by the property $A_{22}^{(1)}=-A_{11}^{(1)}$ \cite{cohentannoudjiAPI} (with all the other coefficients for $q\neq 1$ vanishing). Using QRT, we obtain
\begin{multline}
    \langle \left[ \mathcal{S}_+^{11(q)}(t+\tau) + \mathcal{S}_+^{22(q)}(t+\tau) \right] \mathcal{S}_-(t) \rangle \\ = \ee^{\ii q \omega_L \tau} \left[ \left( C_1^{(q)} + D_1^{(q)} \ee^{-\Gamma_{\mathrm{pop}}\tau} \right) \langle \mathcal{S}_+^{11(q)}(t) \mathcal{S}_-(t) \rangle + \left( C_2^{(q)} + D_2^{(q)} \ee^{-\Gamma_{\mathrm{pop}}\tau} \right) \langle \mathcal{S}_+^{22(q)}(t) \mathcal{S}_-(t) \rangle \right] .
\end{multline}
The expectation values appearing in the above expression can be evaluated as in the case of sidebands, using \eqref{eq:Sijq} and \eqref{eq:Spm},
\begin{align}
    \langle \mathcal{S}_+^{11(q)} (t) \mathcal{S}_-(t) \rangle & = A_{11}^{(q)} \sum_{j\in\{1,2\}} \sum_{\ell} \bigl( A_{j1}^{(\ell)} \bigr)^* \sigma_{j1}^{(q-\ell)} (t) = A_{11}^{(q)} \sum_{\ell} \bigl( A_{11}^{(\ell)} \bigr)^* \Pi_1^{\mathrm{st}} \ee^{\ii(q-\ell)\omega_L t} + f_{11}^{(q)} (t) , \\
    \langle \mathcal{S}_+^{22(q)} (t) \mathcal{S}_-(t) \rangle & = A_{22}^{(q)} \sum_{j\in\{1,2\}} \sum_{\ell} \bigl( A_{j2}^{(\ell)} \bigr)^* \sigma_{j2}^{(q-\ell)} (t) = A_{22}^{(q)} \sum_{\ell} \bigl( A_{22}^{(\ell)} \bigr)^* \Pi_2^{\mathrm{st}} \ee^{\ii(q-\ell)\omega_L t} + f_{22}^{(q)} (t) , 
\end{align}
where $f_{jj}^{(q)} (t)$ contain transient contributions in $t$. Collecting the above results yields
\begin{align}\label{eq:Jc}
    \mathcal{J}_{c}^{(q)}(\omega) & = \frac{\Gamma(\omega)}{\pi} \left[ \mathcal{A}^{(q)} \, \mathrm{Re} \int_0^{\infty} \dd \tau \, \ee^{-\ii(\omega-q\omega_L) \tau } + \mathcal{B}^{(q)} \, \mathrm{Re} \int_0^{\infty} \dd \tau \, \ee^{-\ii(\omega-q\omega_L ) \tau - \Gamma_{\mathrm{pop}}\tau} \right] \nonumber \\
    & = \Gamma(q\omega_L) \mathcal{A}^{(q)} \delta(\omega-q\omega_L) + \Gamma(\omega) \mathcal{B}^{(q)} \frac{1}{\pi} \frac{\Gamma_{\mathrm{pop}}}{ [\omega - (q\omega_L+\Omega)]^2 + \Gamma_{\mathrm{pop}}^2} ,
\end{align}
with
\begin{align}
    \mathcal{A}^{(q)} & = C_1^{q} \bigl\lvert A_{11}^{(q)} \bigr\rvert^2 \Pi_1^{\mathrm{st}} + C_2^{q} \bigl\lvert A_{22}^{(q)} \bigr\rvert^2 \Pi_2^{\mathrm{st}} = \left\lvert A_{11}^{(q)} \Pi_1^{\mathrm{st}} + A_{22}^{(q)} \Pi_2^{\mathrm{st}} \right\rvert^2 , \\ 
    \mathcal{B}^{(q)} & = D_1^{q} \bigl\lvert A_{11}^{(q)} \bigr\rvert^2 \Pi_1^{\mathrm{st}} + D_2^{q} \bigl\lvert A_{22}^{(q)} \bigr\rvert^2 \Pi_2^{\mathrm{st}} = \left\lvert A_{11}^{(q)} - A_{22}^{(q)} \right\rvert^2 \Pi_1^{\mathrm{st}} \Pi_2^{\mathrm{st}} .
\end{align}
After approximating $\Gamma(\omega)\simeq\Gamma(q\omega_L)$, up to $O(\Gamma_{\mathrm{pop}})$ corrections, one can express the spectral density of the $q$-th central line as
\begin{multline}
    \mathcal{J}_{c}^{(q)}(\omega) = \left[ \Gamma_{11}^{(q)} \left( \Pi_1^{\mathrm{st}} \right)^2 + \Gamma_{22}^{(q)} \left( \Pi_2^{\mathrm{st}} \right)^2 + 2\,\mathrm{Re} K_{12}^{(q)} \Pi_1^{\mathrm{st}} \Pi_2^{\mathrm{st}} \right] \delta(\omega - q\omega_L) \\
    + \left( \Gamma_{11}^{(q)} + \Gamma_{22}^{(q)} - 2\,\mathrm{Re} K_{12}^{(q)} \right) \Pi_1^{\mathrm{st}} \Pi_2^{\mathrm{st}} \frac{1}{\pi} \frac{\Gamma_{\mathrm{pop}}}{( \omega - q\omega_L )^2 + \Gamma_{\mathrm{pop}}^2 } .
\end{multline}
It is evident that frequency integration, useful to determine the line weight, cancels the contributions that depend on $K_{12}^{(q)}$.

\section{Perturbative correction of RWA eigenstates}

If the composition of the dressed eigenstates $\{\ket{1(N)},\ket{2(N)}\}$ in terms of the bare states $\ket{e,N}$ and $\ket{g,N}$ is expressed by the amplitudes
\begin{equation}\label{eq:alphabeta}
    \alpha_j^{(p)} = \langle g,N+1-p | j(N) \rangle,  \qquad \beta_j^{(p)} = \langle e,N-p | j(N) \rangle ,
\end{equation}
then the coefficients in Eq.~\eqref{eq:Aq}, which are crucial to determine the features of a specific model, read
\begin{equation}
    A_{ij}^{(q)} = \sum_p \bigl( \beta_i^{(p+q-1)} \bigr)^* \alpha_j^{(p)} .
\end{equation}
In the case of the standard RWA dressed atom, the eigenstates have the form
\begin{equation}\label{eq:j0}
    \ket{1(N)_0} = \sin\theta \ket{g,N+1} + \cos\theta \ket{e,N}, \qquad \ket{2(N)_0} = \cos\theta \ket{g,N+1} - \sin\theta \ket{e,N},
\end{equation}
implying that the only non-vanishing amplitudes \eqref{eq:alphabeta} are those with $p=0$, and the only non-vanishing coefficients in $\mathcal{S}_+$ are those with $q=1$:
\begin{equation}\label{eq:A1}
    A_{11}^{(1)} = \cos\theta \sin\theta, \quad A_{22}^{(1)} = -\cos\theta \sin\theta, \quad A_{12}^{(1)} = \cos^2\theta, \quad A_{21}^{(1)} = - \sin^2\theta .
\end{equation}
Therefore, only the coefficients
\begin{equation}
    A_{ij}^{(0)} = \bigl( \beta_i^{(0)} \bigr)^* \alpha_j^{(0)} + \sum_{p\neq 0} \bigl( \beta_i^{(p)} \bigr)^* \alpha_j^{(p)}
\end{equation}
contain terms of $O(1)$ in perturbations of the RWA Hamiltonian. Incidentally, notice that the quantity $K_{12}^{(1)}$, as defined in Eq.~\eqref{eq:K12}, is negative, because $A_{11}^{(1)}$ and $A_{22}^{(1)}$ have opposite sign. This is ultimately due to the minus sign appearing in the transformation \eqref{eq:j0}, that connects two orthonormal bases.

Let us consider the case in which the Hamiltonian 
\begin{equation}
    H_{\mathrm{AL}} = H_{\mathrm{RW}} + H_{\mathrm{AS}} + H_{\mathrm{CR}}
\end{equation}
contains an ``unperturbed'' term 
\begin{equation}
    H_{\mathrm{RW}} = \hbar\omega_0 \ket{e}\bra{e}\otimes \openone_L + \openone_A\sum_N N\hbar\omega_L \ket{N}\bra{N} + \frac{\hbar\Omega_{\mathrm{R}}}{2} \sum_N \left( \ket{e,N}\bra{g,N+1} + \ket{g,N+1}\bra{e,N} \right) ,
\end{equation}
whose eigenstates are of the form \eqref{eq:j0}, with 
\begin{equation}
    \tan(2\theta) = \frac{\Omega_{\mathrm{R}}}{\omega_0-\omega_L} ,
\end{equation}
and
\begin{align}
    H_{\mathrm{AS}} & = \hbar\Omega_{\mathrm{AS}} \sum_N \left( \ket{e,N}\bra{e,N+1} + \ket{e,N+1}\bra{e,N} \right) , \\
    H_{\mathrm{CR}} & = \frac{\hbar\Omega_{\mathrm{R}}}{2} \sum_N \left( \ket{e,N+1}\bra{g,N} + \ket{g,N}\bra{e,N+1} \right) ,
\end{align}
perturbation terms depending on a permanent dipole moment of the atomic state $\ket{e}$ and on the relevance of counter-rotating terms, respectively. Since $H_{\mathrm{AS}}$ and $H_{\mathrm{CR}}$ do not couple states inside the same doublet, the relevant conditions of perturbation theory applicability are $\Omega_{\mathrm{AS}}\ll\omega_L$ and $\Omega_{\mathrm{R}}\ll\omega_L$. Moreover, to ensure that RWA is a good zeroth-order approximation, the condition 
\begin{equation}\label{eq:RWA}
    \Omega = \sqrt{\Omega_{\mathrm{R}}^2 + (\omega_0-\omega_L)^2}\ll\omega_L ,
\end{equation}
stronger than $\Omega_{\mathrm{R}}\ll\omega_L$, must hold as well, and $\Omega/\omega_L$ will thus be treated as a small parameter.

Applying perturbation theory at the first order, the RWA eigenstates $\ket{j(N)}$ hybridize with $\ket{j'(N+p)}$ with $p=-1,+1$, due to $H_{\mathrm{AS}}$, and $p=-2,+2$, due to $H_{\mathrm{CR}}$. The resulting eigenstates read
\begin{align}
    \ket{1(N)} = \ket{1(N)_0} & + \frac{\Omega_\mathrm{AS}\Omega}{\omega_L(\omega_L+\Omega)} \cos^2\theta \sin\theta \ket{g,N} + \frac{\Omega_\mathrm{AS}}{\omega_L}\cos\theta \left( 1 - \frac{\Omega\sin^2\theta}{\omega_L+\Omega} \right) \ket{e,N-1} \nonumber \\ 
    & + \frac{\Omega_\mathrm{AS}\Omega}{\omega_L(\omega_L-\Omega)} \cos^2\theta \sin\theta \ket{g,N+2} - \frac{\Omega_\mathrm{AS}}{\omega_L}\cos\theta \left( 1 + \frac{\Omega\sin^2\theta}{\omega_L-\Omega} \right) \ket{e,N+1} \nonumber \\
    & + \frac{\Omega_{\mathrm{R}}}{4\omega_L} \frac{\Omega \sin^2\theta\cos\theta}{2\omega_L-\Omega} \ket{g,N+3} - \frac{\Omega_{\mathrm{R}}}{4\omega_L} \sin\theta \left( 1 + \frac{\Omega \sin^2\theta}{2\omega_L-\Omega} \right) \ket{e,N+2} \nonumber \\
    & + \frac{\Omega_{\mathrm{R}}}{4\omega_L} \cos\theta \left( 1 - \frac{\Omega \cos^2\theta}{2\omega_L+\Omega} \right) \ket{g,N-1} + \frac{\Omega_{\mathrm{R}}}{4\omega_L} \frac{\Omega \sin\theta\cos^2\theta}{2\omega_L+\Omega} \ket{e,N-2}
\end{align}
and
\begin{align}
    \ket{2(N)} = \ket{2(N)_0} & - \frac{\Omega_\mathrm{AS}\Omega}{\omega_L(\omega_L-\Omega)} \cos\theta \sin^2\theta \ket{g,N} - \frac{\Omega_\mathrm{AS}}{\omega_L}\sin\theta \left( 1 + \frac{\Omega\cos^2\theta}{\omega_L-\Omega} \right) \ket{e,N-1} \nonumber \\ 
    & - \frac{\Omega_\mathrm{AS}\Omega}{\omega_L(\omega_L+\Omega)} \cos\theta \sin^2\theta \ket{g,N+2} + \frac{\Omega_\mathrm{AS}}{\omega_L}\sin\theta \left( 1 - \frac{\Omega\cos^2\theta}{\omega_L+\Omega} \right) \ket{e,N+1} \nonumber \\
    & + \frac{\Omega_{\mathrm{R}}}{4\omega_L} \frac{\Omega \sin\theta\cos^2\theta}{2\omega_L+\Omega} \ket{g,N+3} - \frac{\Omega_{\mathrm{R}}}{4\omega_L} \cos\theta \left( 1 - \frac{\Omega \cos^2\theta}{2\omega_L+\Omega} \right) \ket{e,N+2} \nonumber \\
    & - \frac{\Omega_{\mathrm{R}}}{4\omega_L} \cos\theta \left( 1 + \frac{\Omega \cos^2\theta}{2\omega_L-\Omega} \right) \ket{g,N-1} - \frac{\Omega_{\mathrm{R}}}{4\omega_L} \frac{\Omega \sin^2\theta\cos\theta}{2\omega_L-\Omega} \ket{e,N-2} .
\end{align}
The lowest-order contribution to the coefficients \eqref{eq:Aq} comes from the terms
\begin{equation}
    A_{ij}^{(q)} \simeq \bigl( \beta_i^{(q-1)} \bigr)^* \alpha_j^{(0)} + \bigl( \beta_i^{(0)} \bigr)^* \alpha_j^{(1-q)} + O \left( \frac{\Omega_{\mathrm{AS}}}{\omega_L} \right)^2 + O \left( \frac{\Omega_{\mathrm{R}}}{\omega_L} \right)^2 .
\end{equation}
with corrections of order $(\Omega_{\mathrm{AS}}/\omega_L)^2$ and $(\Omega_{\mathrm{R}}/\omega_L)^2$. Therefore, the coefficient determining the low-frequency line strength reads
\begin{equation}
    A_{12}^{(0)} = - \frac{\Omega_{\mathrm{AS}}}{\omega_L} \cos^2\theta \left( 1 + 2 \frac{\Omega \sin^2\theta}{\omega_L-\Omega} \right) + O \left[ \left(\frac{\Omega_{\mathrm{AS}}}{\omega_L} \right)^2  \right] = - \frac{\Omega_{\mathrm{AS}}}{\omega_L} \cos^2\theta + O \left(\frac{\Omega_{\mathrm{AS}}\Omega}{\omega_L^2} \right) ,
\end{equation}
while, within the same approximation order, the coefficients $A_{ij}^{(2)}$ that determine the emission triplet around $2\omega_L$ read
\begin{equation}
    A_{11}^{(2)} \simeq \frac{\Omega_{\mathrm{AS}}}{\omega_L} \cos\theta\sin\theta , \quad A_{22}^{(2)} \simeq - \frac{\Omega_{\mathrm{AS}}}{\omega_L} \cos\theta\sin\theta , \quad A_{12}^{(2)} \simeq \frac{\Omega_{\mathrm{AS}}}{\omega_L} \cos^2\theta , \quad A_{21}^{(2)} \simeq - \frac{\Omega_{\mathrm{AS}}}{\omega_L} \sin^2\theta .
\end{equation}
Notice the proportionality to the unperturbed coefficients $A_{ij}^{(1)}$ in Eq.~\eqref{eq:A1}, implying that the relative weight of the low-energy transition with respect to the $q=1$ and the $q=2$ triplets differs only by a factor $(\Omega_{\mathrm{AS}}/\omega_L)^2$ and by the different ratio of form factors. In the case of $q=3$, one can verify that the dominant terms
\begin{equation}
    A_{ij}^{(3)} \simeq \langle i(N) | e,N-2 \rangle \langle g,N+1 | j(N) \rangle + \langle i(N) | e,N \rangle \langle g,N+3 | j(N) \rangle = O \left(\frac{\Omega_{\mathrm{R}}\Omega}{\omega_L^2} \right)
\end{equation}
are suppressed by an additional order $\Omega/\omega_L)$ besides $\Omega_{\mathrm{R}}/\omega_L$, leading to the considerations reported in the main text.

\begin{figure}
    \centering
    \subfigure[$\Omega_{\mathrm{R}}=0.1\omega_L$]{\includegraphics[width=0.32\textwidth]{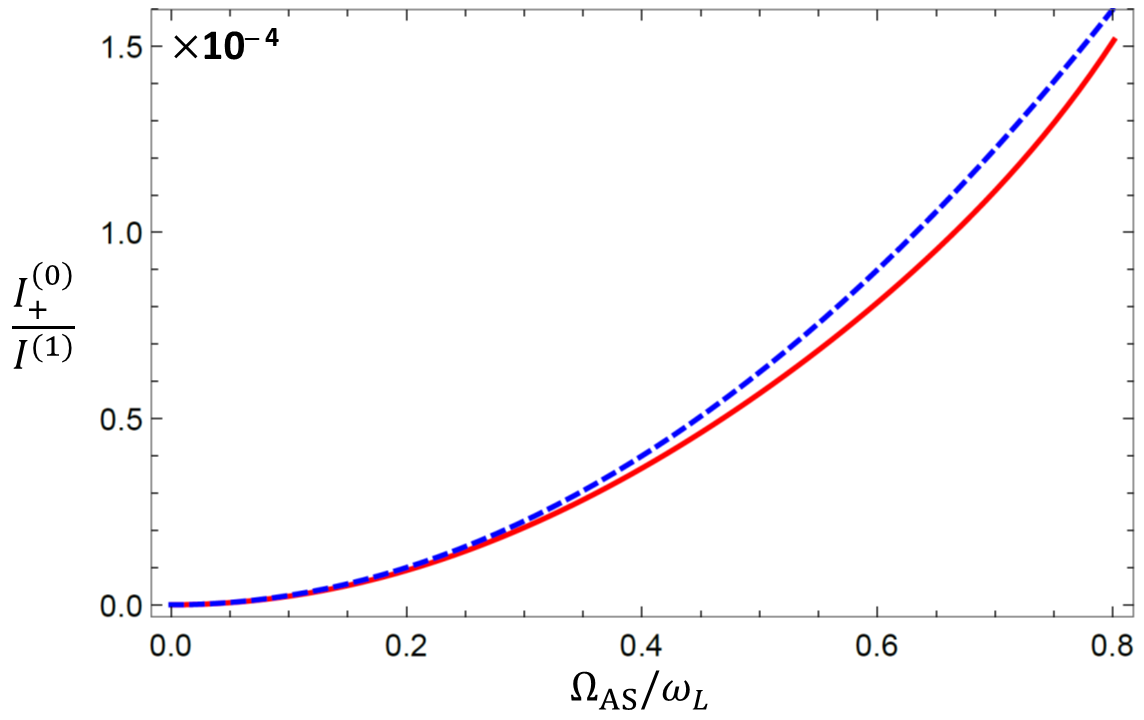}}
    \subfigure[$\Omega_{\mathrm{R}}=0.2\omega_L$]{\includegraphics[width=0.32\textwidth]{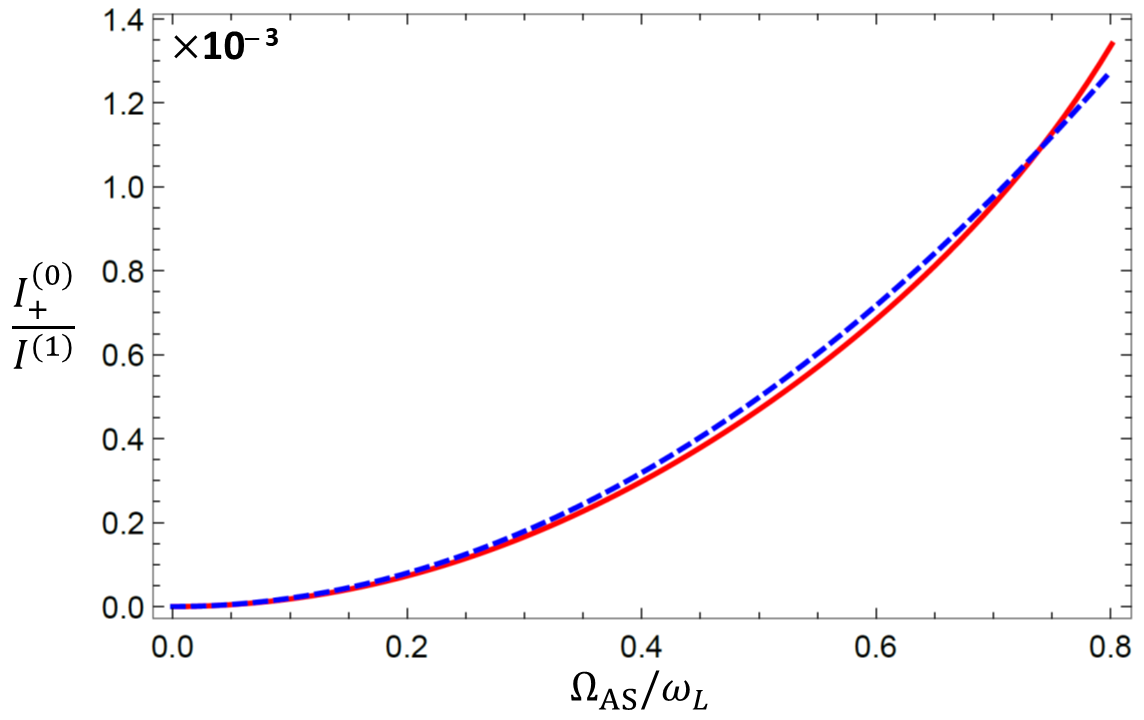}}
    \subfigure[$\Omega_{\mathrm{R}}=0.3\omega_L$]{\includegraphics[width=0.32\textwidth]{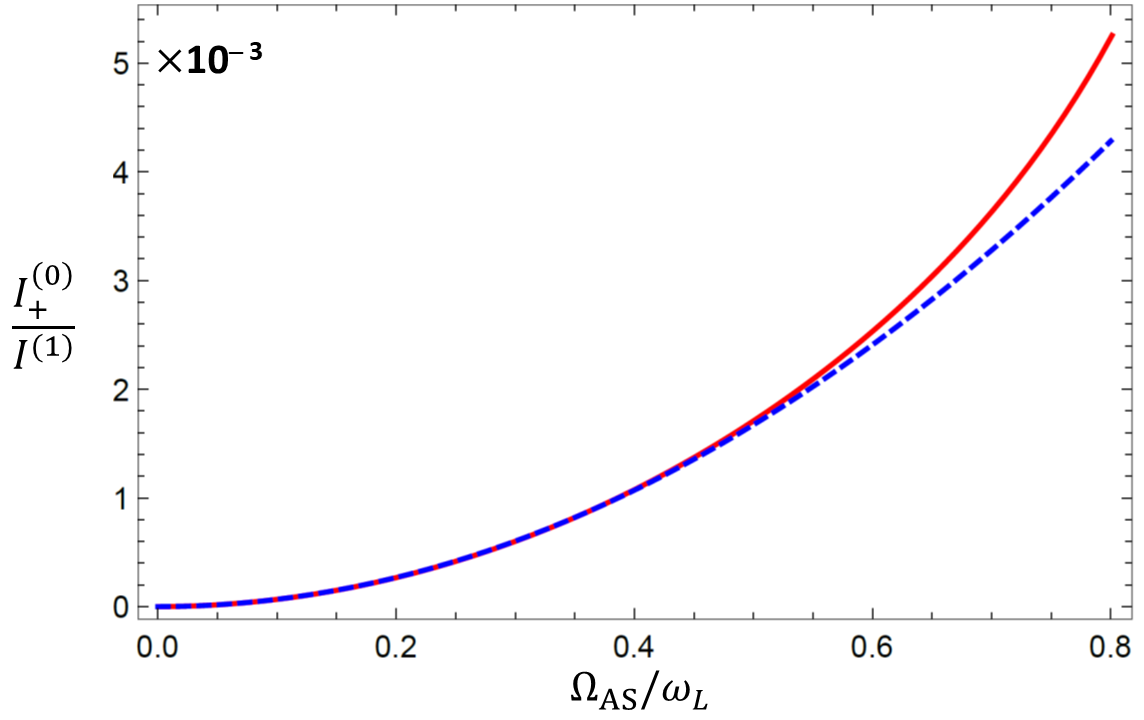}}
    \caption{Relative weight of the low-frequency emission compared to the Mollow triplet, evaluated in resonance conditions $\omega_0=\omega_L$ (solid red lines). The quadratic approximation determined by the behavior around $\Omega_{\mathrm{AS}}=0$ is reported for comparison (dashed blue lines). The results are obtained under the assumption that the form factor $\Gamma(\omega)$ of the atomic transition behaves like $\omega^3$ for the transition frequencies involved in the processes.}
    \label{fig:emission}
\end{figure}

\begin{figure}
    \centering
    \includegraphics[width=0.5\textwidth]{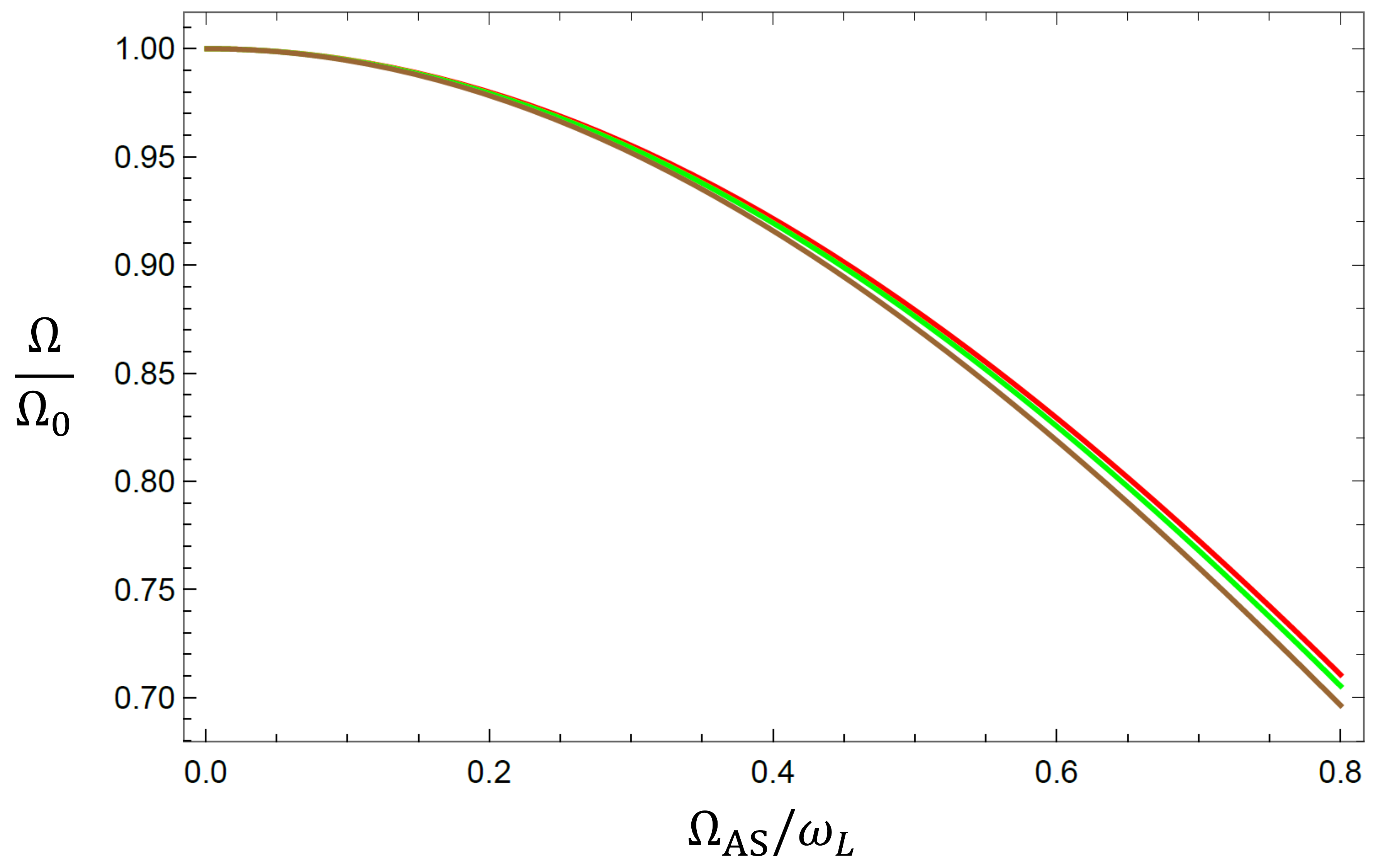}
    \caption{Comparison between the low-frequency gap $\Omega=[E_1(N)-E_2(N)]/\hbar$ obtained with varying $\Omega_{\mathrm{AS}}$ and the value $\Omega_0$ obtained for $\Omega_{\mathrm{AS}}=0$. The reported curves are referred to resonance cases ($\omega_0=\omega_L$) with $\Omega_{\mathrm{R}}=0.1\omega_L$ (red line), $\Omega_{\mathrm{R}}=0.2\omega_L$ (green line), and $\Omega_{\mathrm{R}}=0.3\omega_L$ (brown line).}
    \label{fig:gap}
\end{figure}

A non-perturbative numerical diagonalization of the full Hamiltonian $H_{\mathrm{AL}}$, truncated to 50 levels, provides the results in Figure~\ref{fig:emission}, which report the relative weight, in terms of photon number, of the low-frequency emission compared to the Mollow triplet, as a function of $\Omega_{\mathrm{AS}}$, in resonance conditions for $\Omega_{\mathrm{R}}=0.1\omega_L,0.2\omega_L,0.3\omega_L$. Interestingly, as shown in Figure~\ref{fig:gap}, increasing the permanent dipole while keeping $\Omega_{\mathrm{R}}$ fixed determines a further reduction of the lowest transition frequency in the system.

\twocolumngrid

\end{document}